\documentclass[11pt]{article}
\usepackage[margin=2cm]{geometry}
\usepackage{comment}
\usepackage{amsmath,amssymb,extarrows,mathtools,graphicx,subfigure,setspace}
\usepackage{cite}
\usepackage{slashed}
\usepackage{color}
\makeatother

\numberwithin{equation}{section}

\def\bid{\hbox{1\hspace{-0.04in}I}} %blackboard bold 1
\def\wt{\widetilde}
\newcommand{\be}{\begin{equation}}
\newcommand{\bea}{\begin{eqnarray}}
\newcommand{\eea}{\end{eqnarray}}
\newcommand{\ba}{\begin{array}}
\newcommand{\ea}{\end{array}}
\newcommand{\ee}{\end{equation}}

\def\ba{\begin{array}}
\def\ea{\end{array}}
\def\be{\begin{equation}}
\def\ee{\end{equation}}

\def\-1{^{-1}}

\begin{document}
\onehalfspacing
\noindent
\begin{titlepage}
\hfill
%\vbox{
%    \halign{#\hfil         \cr
%           IPM/P-2013/nnn \cr
%                      } % end of \halign
%      }  % end of \vbox
\vspace*{20mm}
\begin{center}
{\Large {\bf Poisson Lie symmetry and D-branes in  WZW model\\  on the Heisenberg Lie group $H_4$ }\\
}

\vspace*{15mm} \vspace*{1mm}A. Eghbali\footnote{a.eghbali@azaruniv.edu} and {{A. Rezaei-Aghdam\footnote{Corresponding author: rezaei-a@azaruniv.edu} }}
\\

{\it  Department of Physics, Faculty of Basic Sciences,\\
%P.O. Box 19395-5531, Tehran, Iran \\ }
%$^b${\it
 Azarbaijan Shahid Madani University, 53714-161, Tabriz, Iran \\ }

\vspace*{.4cm}

\end{center}
\begin{abstract}
We show that the  WZW model on the Heisenberg Lie group $H_4$ has Poisson-Lie symmetry
only when the dual Lie group is  ${ A}_2 \oplus 2{ A}_1$. In this way, we construct the mutual T-dual sigma models  on  Drinfel'd double generated by the
Heisenberg Lie group $H_4$ and its dual pair, ${ A}_2 \oplus 2{ A}_1$,  as the target space  in such a way that the original model
is the same as the $H_4$ WZW model.
Furthermore, we show that the dual model is conformal  up to two-loop order. Finally, we discuss  $D$-branes and the worldsheet boundary conditions defined by a gluing matrix  on the $H_4$ WZW model.
Using the duality map obtained from the canonical transformation description of the Poisson-Lie T-duality transformations for the gluing matrix which locally defines the properties of the  $D$-brane,  we find two different cases of the gluing matrices for the WZW model based  on the Heisenberg Lie group $H_4$ and its dual model.
\end{abstract}
\end{titlepage}
%%%%%%%%%%%%%%%%%%%%%%%%%%%%%%%%%%%%%%%%%%%%%%%%%%%%%%%%%%%%%%%%%%%%%%%%%%%%%%
%%%%%%%%%%%%%%%%%%%%%%%%%%%%%%%%%%%%%%%%%%%%%%%%%%%%%%%%%%%%%%%%%%%%%%%%%%%%%%%%%%%%%%%%%%%%%%%%%%%%%%

\section{\large Introduction}

The WZW models on non-semi-simple Lie groups \cite{Witten}, \cite{Kiritsis}, \cite{Sfetsos}, and \cite{Kehagias} play an important role in string theory, since some of them
provide exact string backgrounds that have a target space dimension equal to the integer and irrational
Virasoro central charge of the affine non-semi-simple algebra \cite{Sfetsos} and \cite{rezaei-aghdam}. The first of these
models was based on the group $E_2^c$, a central extension of the two-dimensional
Euclidean group; and the corresponding sigma model describes  string propagation  on a  four-dimensional
space-time in the background of a gravitational plane wave \cite{Witten}. This
construction was subsequently extended to other non-semisimple  Lie groups \cite{Sfetsos} and \cite{Kehagias} in such a way that the
WZW model on the Heisenberg group  with arbitrary dimension was, for the first time, introduced by Kehagias and Meessen \cite{Kehagias}.

On the other hand, the T-duality is a very
important symmetry of string theories, or more generally,
two-dimensional sigma models \cite{Busher} and the Poisson-Lie T-duality \cite{Klim1}, \cite{Klim2} is a generalization of Abelian
and non-Abelian  target space duality (T-duality).
So far, there is one
example for  conformal  sigma models related by Poisson-Lie T-duality
\cite{Klim3} in a way that the duality relates the standard $SL(2,R)$ WZW
model to a constrained sigma model defined on the $SL(2,R)$ group space.
Moreover, we have recently shown  that the WZW models on the Lie supergroups $GL(1|1)$ \cite{ER7} and $(C^3+A)$ \cite{ER8} contain
super Poisson-Lie symmetry such that in this process the dual Lie supergroups are the respective
${B \oplus A \oplus A_{1,1}}_|.i$  and  ${C^3 \oplus {A}_{1,1}}_|.i$. In this paper we show that the  WZW model on the Heisenberg Lie group $H_4$ has Poisson-Lie symmetry
only when the dual Lie group is  ${ A}_2 \oplus 2{ A}_1$.
Furthermore, we show that the dual model is conformal  up to two-loop order and in this manner we obtain the general form of the dilaton field of
the dual model.
We also study the worldsheet boundary conditions for our model and its dual.

The outline of the paper is as follows. In section 2 we  show that the  WZW model on the Heisenberg Lie group $H_4$ has Poisson-Lie symmetry
only when the dual Lie group is  ${ A}_2 \oplus 2{ A}_1$. In section 3, we first construct the Poisson-Lie T-dual sigma models  on
the Drinfel'd double $(H_4~,~ { A}_2 \oplus 2{ A}_1)$ in such a way that we show the original model
is the same as the $H_4$ WZW model. Then, by calculation of the vanishing of the one-loop $B$-functions we obtain the general form of the dilaton field of
the dual model and followed by,
we show that the dual model is conformal up to two-loop order.  Finally, in section 4, we first review the worldsheet boundary conditions
under the Poisson-Lie T-duality and reobtain, in general, the algebraic form of a duality map for the gluing matrix between both the original and dual
models under the Poisson-Lie T-duality transformation. Then, we study
the consequences of the duality transformation of the gluing matrix for the $H_4$ WZW model and its dual model.  Some concluding remarks are given  in the last section.

%%%%%%%%%%%%%%%%%%%%%%%%%%%%%%%%%%%%%%%%%%%%%%%%%%%%%%%%%%%%%%%%%%%%%%%%%%%%%%%%%%%%%%%%%%%%%%%%%%%%%%%%%%%%
\section{\large Poisson-Lie symmetry of the  WZW model on the Heisenberg Lie group $H_4$}

In this section, based on our previous works \cite{ER7} and \cite{ER8},  we will describe a new example of a WZW model containing
Poisson-Lie symmetry. The model is constructed on the Heisenberg Lie group $H_4$, a  non-semi-simple  Lie group of dimension four.  As mentioned in introduction section, the WZW model based on the Heisenberg group $H_4$ was, for the first time, constructed by Kehagias and Meessen \cite{Kehagias}. Here, we first obtain
the WZW model on the Heisenberg group $H_4$ with a new background. Then, we will show that
the model has Poisson-Lie symmetry.
Before proceeding to construct model, let us first introduce  the  Lie algebra $h_4$ of the  Lie group $H_4$ (the oscillator Lie algebra).
The  Lie algebra $h_4$ is generated by the generators $\{N, A_+, A_-, M\}$ with the following non-zero Lie brackets
\begin{eqnarray}\label{1}
[N , A_+]~=~A_+,~~~~~[N , A_-]~=~-A_-,~~~~~[A_- , A_+]~=~M.
\end{eqnarray}
One can show that the  Lie algebra  $h_4$ is isomorphic to the Drinfel'd double of a two-dimensional Lie bialgebra, i.e., $({\cal A}_2 , {\cal I}_2)$ \cite{Hlavaty1}, \cite{sephid} where ${\cal A}_2$ and ${\cal I}_2$ are two-dimensional non-Abelian and Abelian  Lie algebras, respectively.
The isomorphic transformation between the Lie algebras $h_4$ and $({\cal A}_2 , {\cal I}_2)$ is given by
\begin{eqnarray*}
N~=~T_1+\alpha_0 T_4,~~~~~A_+~=~\beta_0 T_2-\alpha_0 \beta_0 T_3,~~~~~A_-~=~\gamma_0 T_4,~~~~~M~=~-\beta_0 \gamma_0 T_3,
\end{eqnarray*}
where $\{T_1,\cdots,T_4\}$ are generators of the  Lie algebra of the  Drinfel'd double  $({\cal A}_2 , {\cal I}_2)$ and $\alpha_0 \in \Re;~\beta_0, \gamma_0 \in \Re-\{0\}$.

Let us now turn into the construction of our model. In general, given a Lie algebra with generators $X_a$ and structure constants $f_{ab}^{~~c}$, to define a WZW model,
one needs a non-degenerate ad-invariant symmetric bilinear form $\Omega_{ab} = <X_a~ , ~X_b >$  on Lie algebra ${\cal G}$
such that it satisfies the following relation \cite{Witten}
\begin{eqnarray}\label{2}
f_{ab}^{\;\;d} \;\Omega_{dc}+ f_{ac}^{\;\;d} \;\Omega_{db}\;=\;0.
\end{eqnarray}
Using the commutation relations (\ref{1}), one can obtain the non-degenerate ad-invariant bilinear form $\Omega_{ab}$ on the  Lie algebra $h_4$ as
{\small \begin{eqnarray}\label{3}
\Omega_{ab}~=~\left( \begin{tabular}{cccc}
              $0$ & $0$ & $0$ & $-\kappa_0$\\
              $0$ & $0$ & $\kappa_0$ & $0$ \\
              $0$ & $\kappa_0$ & $0$ & $0$ \\
              $-\kappa_0$ & $0$ & $0$ & $0$\\
                \end{tabular} \right),
\end{eqnarray}}
where $\kappa_0$ is a non-zero real constant. In general, we know that the WZW model based on a Lie group $G$  is defined on a Riemannian surface $\Sigma$
as a worldsheet by the following action \cite{Witten}
\begin{eqnarray}\label{4}
S_{_{WZW}}(g) &=&  \frac{K}{4\pi} \int_{\Sigma} d\sigma^+ d\sigma^-\;
L^{\hspace{-0.5mm}a}_{+}\;{\Omega}_{ab}\;
L^{\hspace{-0.5mm}b}_{-}
+\frac{K}{24\pi} \int_{B} d^3 \sigma~
\varepsilon^{ \gamma \alpha \beta}
L^{\hspace{-0.5mm}a}_{\gamma}
\;{\Omega}_{ad}\;L^{\hspace{-0.5mm}b}_{\alpha}
\;f_{bc}^{~~d}~ L^{\hspace{-0.5mm}c}_{\beta},
\end{eqnarray}
where the components of the left-invariant one-forms $L^{\hspace{-0.5mm}a}_{\alpha}$'s are defined via $g^{-1}
\partial_{\alpha}g\;=\;L^{\hspace{-0.5mm}a}_{\alpha}~ X_{a}$, in which  $g:  \Sigma \rightarrow G$ is an element of Lie group $G$. Here
{\small $B$} is a three-manifold bounded by worldsheet $\Sigma$ and
$\sigma^{\pm} \;=\;
\frac{1}{\sqrt{2}}(\tau \pm \sigma)$ are the standard light-cone variables on the worldsheet.
To calculate the
$L^{\hspace{-0.5mm}a}_{\alpha}$'s  we parameterise the  corresponding  Lie group $H_4$ with coordinates $x^{\mu} =\{x, y, u, v\}$
so that its elements can
be written as
\begin{eqnarray}\label{5}
g\;=\; e^{v X_4} ~ e^{u X_3} ~ e^{x X_1}~ e^{y X_2},
\end{eqnarray}
where  we have introduced the new generators $\{X_1, X_2, X_3, X_4\}$ instead of $\{N, A_+, A_-, M\}$, respectively.
We then obtain
\begin{eqnarray}\label{6}
L^{\hspace{-0.5mm}1}_{\pm}=\partial_{\pm} x,~~~~~L^{\hspace{-0.5mm}2}_{\pm}=y \partial_{\pm} x + \partial_{\pm} y,~~~~~
L^{\hspace{-0.5mm}3}_{\pm}=e^{x}~\partial_{\pm} u,~~~~~L^{\hspace{-0.5mm}4}_{\pm}=ye^{x}~ \partial_{\pm} u + \partial_{\pm} v.
\end{eqnarray}
Hence using relations (\ref{1}), (\ref{3}) and (\ref{6}) and some algebraic calculations, the WZW action on the $H_4$ Lie group is worked out to be of the form
\begin{eqnarray}
S_{_{WZW}}(g) &=&\frac{\kappa_0 K}{4 \pi} \int d\sigma^+d\sigma^-\Big\{-\partial_+x \partial_-v -\partial_+v \partial_-x
+e^x\Big(\partial_+y \partial_-u+\partial_+u \partial_-y\nonumber\\
&&~~~~~~~~~~~~~~~~~~~~~+y\partial_+u \partial_-x- y\partial_+x \partial_-u\Big) \Big\}.\label{7}
\end{eqnarray}
As we know the non-linear sigma model  for a bosonic string propagating in a $d$-dimensional  space-time with the metric $g_{{\mu\nu}}$,
the anti-symmetric tensor field $b_{\mu\nu}$ and the dilaton fileld $\Phi$ is given by\footnote{The dimensional coupling constant $\alpha'$ turns out to be the inverse string tension.}
\begin{eqnarray}\label{666.1}
S~=~\frac{1}{2 \pi \alpha'}\int\!d\tau  d\sigma\;\sqrt{-h} \left\{  \frac{1}{2} \Big(h^{\alpha \beta}g_{_{\mu\nu}}
+\epsilon^{\alpha \beta} b_{_{\mu\nu}}\Big)~\partial_{\alpha}x{^{^\mu}}
\partial_{\beta}x^{^{\nu}} + \frac{1}{4} \alpha' \Phi(x^\mu) ~ R^{^{(h)}}\right\},
\end{eqnarray}
where $h_{\alpha \beta}$  and  $\epsilon^{\alpha \beta}$ are  the worldsheet metric with $R^{^{(h)}}$ the corresponding  worldsheet curvature scalar
and anti-symmetric tensor on the worldsheet, respectively, such that
$h:= \det {h_{\alpha \beta}}$ and the indices $\alpha, \beta=\tau, \sigma$. Note that here we consider $\Phi(x^\mu) =0$. The model (\ref{666.1}) is invariant under worldsheet
reparametrisation, therefore
this symmetry allows us to switch to light-cone coordinates on the worldsheet; consequently,  in the absence of the dilaton we have \cite{Klim1}
\begin{eqnarray}\label{8}
S =\frac{1}{2} \int d\sigma^+ d\sigma^-~  \big({g}_{\mu \nu}+{b}_{\mu \nu}\big)~\partial_+x^{\mu} \partial_-x^{\nu}.
\end{eqnarray}
Here the space-time geometry is described by a Lorentz signature metric $g_{\mu \nu}$ and anti-symmetric tensor field $b_{\mu \nu}$, both
of which may depend on the space-time coordinates $x^\mu$. For the action (\ref{7}), the corresponding
space-time  metric and the anti-symmetric tensor field  are, respectively, given by
\begin{eqnarray}\label{666}
ds^2&=&-2 dx dv + 2 e^x~ dy du,\nonumber\\
b&=&-ye^x~ dx \wedge du.
\end{eqnarray}
Thus, by identifying the action (\ref{7}) with the sigma model
of the form (\ref{8}) we can read off the background matrix
${\cal E}_{\mu \nu}={g}_{\mu \nu}+{b}_{\mu \nu}$ in the coordinate base $\{dx, dy, du,dv\}$ as
\begin{eqnarray}\label{9}
{\cal E}_{\mu \nu}~=~\left( \begin{tabular}{cccc}
              $0$ & $0$ & $-y e^x$ & $-1$\\
              $0$ & $0$ & $e^x$ & $0$ \\
              $y e^x$ & $e^x$ & $0$ & $0$ \\
              $-1$ & $0$ & $0$ & $0$\\
                \end{tabular} \right).
\end{eqnarray}
The sigma model (\ref{8}) has Poisson-Lie symmetry with respect to the dual Lie group
 ${\tilde G}$ (with the same dimension $G$) when background matrix satisfies in
 the following relation \cite{Klim1}, \cite{Klim2}
\begin{eqnarray}\label{10}
{\cal L}_{V_a}({\cal E}_{\mu \nu})~=~
 {\cal E}_{\mu \rho} ~{V_b}^{~\rho}\;{{\tilde
f}^{cb}}_{~~a}\;{V_c}^{~\lambda}\;{\cal E}_{\lambda \nu},
\end{eqnarray}
where ${\cal L}_{V_a}$ stands for the Lie derivative corresponding to the left invariant vector field ${V_a}$
and ${{\tilde
f}^{cb}}_{~~a}$ are the structure constants of the dual Lie algebra ${\tilde {\cal G}}$.
In the following we will show that the WZW model on the $H_4$ Lie group has Poisson-Lie symmetry. To this end,
we need  the left invariant vector fields on the $H_4$. Utilizing relation (\ref{6})
in the equation ${{V_a}}^{~\mu}~{L_{\mu}}^{~b}=\delta_a^{~b}$
the ${V_a}$'s are obtained to be
\begin{eqnarray}\label{11}
{V}_{1}=\frac{\partial}{\partial x}-y \frac{\partial}{\partial y},~~~~~{V}_{2}= \frac{\partial}{\partial y},~~~~~
{V}_{3}=e^{-x}~\frac{\partial}{\partial u}-y \frac{\partial}{\partial v},~~~~~{V}_{4}=\frac{\partial}{\partial v}.
\end{eqnarray}
Now, by substituting relations (\ref{9}) and (\ref{11}) on the right hand side of (\ref{10}) and, then, by
direct calculation of Lie derivative of ${\cal E}_{\mu \nu}$ with respect to $V_a$, one can obtain the structure constants of the  dual Lie algebra to the  Lie
algebra $h_4$ in such a way that only non-zero commutation relation of the dual pair is
\begin{eqnarray}\label{12}
[{\tilde X}^2 , {\tilde X}^4]~=~{\tilde X}^2.
\end{eqnarray}
The Lie algebra deduced in this process is a four-dimensional decomposable Lie algebra. In the classification of four-dimensional Lie algebras
\cite{patera} it has denoted  by ${\cal A}_2 \oplus 2{\cal A}_1$ where
${\cal A}_1$ is one-dimensional Lie algebra. Note that both sets of generators (\ref{1}) and (\ref{12}) are maximally isotropic with respect
to the non-degenerate invariant symmetric bilinear form defined by the brackets \cite{Klim2}
\begin{eqnarray*}
<X_a~,~X_b>~=~<{\tilde X}^a~ ,~ {\tilde X}^b>~=~0,~~~~~~~~~<X_a~,~{\tilde X}^b>~=~{\delta_a}^{~b}.
\end{eqnarray*}
Nevertheless, the $(h_4~,~ {\cal A}_2 \oplus 2{\cal A}_1)$ as a Lie bialgebra satisfies mixed Jacobi identities.
Having a Drinfel'd double which is simply a Lie group $D$, we can construct the Poisson-Lie symmetric sigma models on it.
So we will, first, form the Drinfel'd double generated by the  Lie algebra $h_4$ and its
dual pair ${\cal A}_2 \oplus 2{\cal A}_1$.
The Manin triple\footnote{The Lie algebra ${\cal D}$ provided with non-degenerate ad-invariant symmetric bilinear form $<.~,~.>$
will be called Drinfel'd double
iff it can be decomposed into a pair of maximally isotropic sub-algebras ${\cal G}$ and  ${\tilde {\cal G}}$ such that
${\cal D} = {\cal G}
\oplus {\tilde {\cal G}}$. The triple $({\cal D} , {\cal G} , {\tilde {\cal G}})$ is called Manin triple.}
$({\cal D} , h_4 , {\cal A}_2 \oplus 2{\cal A}_1)$  possesses eight generators $\{X_1,..., X_4; {\tilde X}^1,..., {\tilde X}^4\}$
so that they obey the following set of non-zero commutation relations
\begin{eqnarray}
{[X_1~,~X_2] }& = & X_2,~~~~~~~~~\;\,[X_1~,~X_3]=-X_3,~~~[X_2~,~X_3]=-X_4,~~~\,[{\tilde X}^2 ~, ~{\tilde X}^4]={\tilde X}^2,\nonumber\\
{[X_1~,~{\tilde X}^2] }& = & -{\tilde X}^2,~~~~~~~\,\,[X_1~,~{\tilde X}^3]={\tilde X}^3,~~~~\,[X_3~,~{\tilde X}^3]=-{\tilde X}^1,~~~
[X_3 ~, ~{\tilde X}^4]=-{\tilde X}^2,\nonumber\\
{[X_2~,~{\tilde X}^2]} & = & X_4+{\tilde X}^1,~~~ [X_2~,~{\tilde X}^4]=-X_2+{\tilde X}^3.~~\label{13}
\end{eqnarray}
In the next section, we shall construct
a pair of Poisson-Lie T-dual sigma models which is associated with the Drinfel'd double $(H_4~,~ { A}_2 \oplus 2{ A}_1)$  and will show that
the original sigma model on the $H_4$ is the same as the WZW model obtained in (\ref{7}).

%%%%%%%%%%%%%%%%%%%%%%%%%%%%%%%%%%%%%%%%%%%%%%%%%%%%%%%%%%%%%%%%%%%%%%%%%%%%%%%%%%%%%%%%%%%
\section{\large Poisson-Lie T-dual sigma models built on the  Drinfel'd double $(H_4~,~ { A}_2 \oplus 2{ A}_1)$ }
As mentioned above,  having  Drinfel'd doubles we can construct the Poisson-Lie T-dual sigma models on them. The construction
of the models has been described in \cite{Klim1} and \cite{Klim2}. The models have target spaces as the Lie groups $G$ and ${\tilde G}$ and are, respectively,  given by the actions
\begin{eqnarray}
S &=& \frac{1}{2} \int d\sigma^+~d\sigma^-~E^+_{ab}(g)~R_{+}^a~ R_{-}^b,\label{14}\\
{\tilde S} &=& \frac{1}{2} \int d\sigma^+~d\sigma^-~{\tilde E}^+{^{ab}}(\tilde g)~({\tilde R}_{+})_a~
({\tilde R}_{-})_b,\label{15}
\end{eqnarray}
where $R_{\pm}^a$ and $({\tilde R}_{\pm})_a$ are  the components of the right-invariant one-forms on the Lie groups $G$ and ${\tilde G}$, respectively, and are defined by
\begin{eqnarray}\label{15.1111}
R_{\pm}^a := (\partial_{\pm}g ~g^{-1})^a =\partial_{\pm}x^\mu~ R_\mu^{~a},~~~~~~
({\tilde R}_{\pm})_a := (\partial_{\pm}{\tilde g} ~{\tilde g}^{-1})_a = \partial_{\pm} {\tilde x}^\mu~ {\tilde R}_{\mu a},
\end{eqnarray}
furthermore, the background fields $E^+(g)$ and ${\tilde E}^+(\tilde g)$ are defined by
\begin{eqnarray}\label{15.1}
E^+(g) = \left(\Pi (g) + (E^+_0)^{-1}(e)\right)^{-1},~~~~{\tilde E}^+(\tilde g) = \left({\tilde \Pi} (\tilde g) + ({\tilde E}^+_0)^{-1}(\tilde e)\right)^{-1},
\end{eqnarray}
in which $E^+_0(e)$ and ${\tilde E}^+_0 (\tilde e)$ are the sigma model constant matrices at the
unit element of $G$ and $\tilde G$, respectively, and are related to
each other in the  following way\cite{Klim2}
\begin{eqnarray}\label{15.12}
E^+_0(e) {\tilde E}^+_0(\tilde e) ~=~{\tilde E}^+_0(\tilde e) E^+_0(e)  ~=~\bid.
\end{eqnarray}
Here  $\Pi(g)$ is a bivector field on the Lie group manifold which gives a Poisson-Lie bracket
on $G$ and is defined as
\begin{eqnarray}\label{15.13}
\Pi^{ij}(g)~=~b^{ik}(g) ({a^{-1}})_k^{~j}(g),
\end{eqnarray}
where $a(g)$ and $b(g)$ are sub-matrices of the adjoint representation of $G$ on the Lie algebra of the  Drinfel'd double \cite{Klim2}.
Analogously ${\tilde \Pi}({\tilde g})$
is a Poisson-Lie bracket on the dual Lie group manifold ${\tilde G}$.

To construct the mutual T-dual sigma models with the Drinfel'd double
$(H_4~,~ { A}_2 \oplus 2{ A}_1)$  whose Lie algebra defined in (\ref{13}), we use the same parameterisation (\ref{5}) for both
the original and dual models\footnote{The parameters $\tilde x$, $\tilde y$, $\tilde u$ and $\tilde v$ are applied for the dual model.}. Using relation (\ref{5}) and (\ref{15.1111}), one can calculate $R_{\pm}^a$'s as
\begin{eqnarray}\label{16.0}
\begin{array}{rclrcl}
R_{\pm}^1 &=& \partial_{\pm} x,
&R_{\pm}^2 &=& e^x~\partial_{\pm} y,\\
R_{\pm}^3 &=& u~\partial_{\pm} x+ \partial_{\pm} u,
&R_{\pm}^4 &=& \partial_{\pm} v + u e^x~\partial_{\pm} y,\\
\end{array}
\end{eqnarray}
and considering relation (\ref{5}) for the corresponding tilted symbols, we obtain
\begin{eqnarray}\label{16.00}
\begin{array}{rclrcl}
({{\tilde R}_{\pm}})_1 &=& \partial_{\pm} {\tilde x},&
({{\tilde R}_{\pm}})_2 &=& e^{-\tilde v}~\partial_{\pm} {\tilde y},\\
({{\tilde R}_{\pm}})_3 &=& \partial_{\pm} {\tilde u},&
({{\tilde R}_{\pm}})_4 &=& \partial_{\pm} {\tilde v}.
\end{array}
\end{eqnarray}
Then, choosing the  constant matrix  at the unit element of $H_4$ as
{\small \begin{eqnarray}\label{16}
E^+_0(e)~=~\left( \begin{tabular}{cccc}
              $0$ & $0$ & $0$ & $1$\\
              $0$ & $0$ & $-1$ & $0$ \\
              $0$ & $-1$ & $0$ & $0$ \\
              $1$ & $0$ & $0$ & $0$\\
                \end{tabular} \right),
\end{eqnarray}}
T-dual sigma models (\ref{14}) and (\ref{15}) are found to be given by
\begin{eqnarray}
S&=&\frac{1}{2} \int d\sigma^+~d\sigma^-\Big\{\partial_+x \partial_-v +\partial_+v \partial_-x
-e^x\Big(\partial_+y \partial_-u+\partial_+u \partial_-y+y\partial_+u \partial_-x\nonumber\\
&&~~~~~~~~~~~~~~~~ ~~~~- y\partial_+x \partial_-u\Big) \Big\},\hspace{1cm}\label{17}\\
{\tilde S}&=&\frac{1}{2} \int d\sigma^+~d\sigma^-\Big\{\partial_+{\tilde x } \partial_-{\tilde v} +\partial_+{\tilde v} \partial_-{\tilde x}
-\partial_+{\tilde u} \partial_-{\tilde y}
+{\tilde u} \partial_+{\tilde v} \partial_-{\tilde y}+{\tilde y} \partial_+{\tilde u} \partial_-{\tilde v}\nonumber\\
&&~~~~~~~~~~~+\frac{e^{-\tilde v}}{e^{-\tilde v}-2}
\Big(\partial_+{\tilde y} \partial_-{\tilde u} + {\tilde u} \partial_+{\tilde y} \partial_-{\tilde v}+ {\tilde y}
\partial_+{\tilde v} \partial_-{\tilde u}+2{\tilde y}{\tilde u} e^{\tilde v}
\partial_+{\tilde v} \partial_-{\tilde v}\Big) \Big\}.\label{18}
\end{eqnarray}
Now by rescaling $\kappa_0$ to $\frac{-2 \pi}{K}$ in action (\ref{7})  one can conclude that action (\ref{17})  is nothing but the WZW action based on the
Lie group $H_4$. Thus, we showed that the Poisson-Lie T-duality relates the
$H_4$ WZW model to a sigma model defined on the dual Lie group  of $H_4$, i.e.,  ${ A}_2 \oplus 2{ A}_1$. It is seen that in this case, the Poisson-Lie T-duality transforms
rather extensive and complicated action  (\ref{18}) to much simpler form such as (\ref{17}).

%%%%%%%%%%%%%%%%%%%%%%%%%%%%%%%%%%%%%%%%%%%%%%%%%%%%%%%%%%%%%%%%%%%%%%%%%%%%%%%%%%%%%%%%%%%
\subsection{\large Conformal invariance of the dual sigma model up to the two-loop $B$-functions}

Consistency of the string theory requires that the action (\ref{666.1})
defines a conformally invariant quantum field theory, and the conditions for conformal
invariance can be interpreted as effective field equations for $g_{\mu \nu}$,  $b_{\mu \nu}$ and $\Phi$ of the string effective  action \cite{{A.Sen},{callan}}.
The conditions for conformal invariance of the sigma model to the order $\alpha'^2$ (at the two-loop level) have been obtained in \cite{c.hull} (see also Ref. \cite{Tseyt}).
It has shown that these  conditions
can be derived from any one of a family of space-time effective actions.

When the $\beta$-functions are trivial, i.e.,  they vanish up to the ambiguities inherent in
their definition, then the theory is rigid scale invariant, i.e.,  the integrated trace
anomaly vanishes. The local scale or conformal invariance needed here requires that the trace anomaly vanishes locally, which requires
the vanishing of certain {\it $B$-functions} \cite{c.hull}.  The  vanishing of two-loop $B$-function gives us the conformal invariance conditions of the sigma model (\ref{666.1}) to the order $\alpha'^2$ \cite{c.hull}. In the following, we use these conditions to show that the dual sigma model (\ref{18}) is  conformally invariant up to two-loop order.

\subsubsection{\it \large Conformal invariance of the dual sigma model up to the one-loop $B$-functions}

The conditions for conformal invariance to hold in two-dimensional sigma model (\ref{666.1}) in the lowest non-trivial approximation are the vanishing of the one-loop $B$-functions. The one-loop $B$-functions are given by \cite{c.hull}
\begin{eqnarray}
B^{^{g}}_{\mu\nu} &=&-\alpha'\Big[R_{{\mu \nu}}-(H^2)_{\mu \nu}+{\nabla}_\mu
{\nabla}_\nu \Phi\Big] + {\cal O}(\alpha'^2),\nonumber\\
B^{^{b}}_{\mu\nu} &=&-\alpha'\Big[-{\nabla}^\lambda H_{{\lambda \mu \nu}} + H_{{\mu \nu}}^{~\;\lambda}  {\nabla}_\lambda\Phi\Big] +{\cal O}(\alpha'^2),\nonumber\\
B^{^{\Phi}} &=&-\alpha'\Big[-\frac{1}{2} {\nabla}^2 \Phi + \frac{1}{2} ({\nabla} \Phi)^2-\frac{1}{3} H^2 \Big] + {\cal O}(\alpha'^2),\label{18.1}
\end{eqnarray}
where  $R_{{\mu \nu}}$ is the  Ricci tensor of the metric $g_{\mu \nu}$,
\begin{eqnarray}
H_{{\mu \nu \rho}} = \frac{1}{2} \big(\partial_{\mu} b_{\nu \rho}+\partial_{\nu} b_{\rho \mu}+\partial_{\rho} b_{\mu \nu}\big),
\end{eqnarray}
is the torsion of the anti-symmetric field $b_{\mu \nu}$, $(H^2)_{\mu \nu} = H_{{\mu \rho \sigma }} H^{{\rho \sigma}}_{~~\nu}$ and $H^2 = H_{{\mu \nu \rho}} H^{{\mu \nu \rho}} $.

Notice that the original model (the model described by action (\ref{17}))
as a WZW model should be conformally invariant. One can find that the only non-zero components of $R_{{\mu \nu}}$
and $H$ are $R_{xx} = -\frac{1}{2}$ and $H_{xyu} = - \frac{e^x}{2}$, respectively. Thus, it is straightforward  to get $R=0=H^2$
and verify that the only non-zero component of $(H^2)_{\mu \nu}$ is $2 H_{xyu} H^{yu}_{\;\;~x} = -\frac{1}{2}$. Consequently, the metric  of this model is flat in the sense that its scalar curvature
vanishes. Employing  the above results in the vanishing of equations (\ref{18.1}) we conclude that dilaton is constant. Nevertheless, by solving  the vanishing of equations
(\ref{18.1}) one can also find a non-constant dilaton as
\begin{eqnarray}
\Phi(x^\mu)~=~\Phi_0 + {\cal C} x,\label{18.2}
\end{eqnarray}
where $\Phi_0$ and ${\cal C}$ are integration constants.

Let us now turn into the dual model. With  regard to action (\ref{18}) the line element of the dual model is
\begin{eqnarray}
d{\tilde s}^2~=~2d{\tilde x} d{\tilde v}+\frac{2}{e^{-\tilde v}-2}\Big[d{\tilde y} d{\tilde u}
+{\tilde u} ({e^{-\tilde v}-1})~ d{\tilde y} d{\tilde v}+{\tilde y} ({e^{-\tilde v}-1})~ d{\tilde u} d{\tilde v}  + {\tilde y} {\tilde u}
~d{\tilde v}^2 \Big].\label{18.3}
\end{eqnarray}
Analogously, for the dual model we find that the only non-zero component of ${\tilde R}_{{\mu \nu}}$
is ${\tilde R}_{{\tilde v}{\tilde v}} = -\frac{1}{2(e^{-\tilde v}-2)^2}\Big(e^{-2\tilde v} -4 e^{-\tilde v}+16 \Big)$; as ${\tilde g}^{{\tilde v}{\tilde v}}=0$, ${\tilde R}=0$.
Therefore, the metric of dual model is also flat in the sense that its scalar curvature
vanishes. As shown in the above, the dual metric has an apparent singularity. This singularity is the coordinate singularity in the metric.
In general relativity \cite{wald}, to investigate the types of singularities one has to study the invariant
characteristics of space-time. To detect the singularities it is sufficient to study only three of them, the
Ricci scalar $R$, $R_{\mu \nu} R^{\mu \nu}$ and the so-called Kretschmann scalar $R_{\mu \nu \sigma\rho } R^{\mu \nu \sigma\rho}$.
For the line element (\ref{18.3}), ${\tilde R}$, ${\tilde R}_{\mu \nu} {\tilde R}^{\mu \nu}$ and the Kretschmann scalar  vanish. Therefore, the singular point
is not an essential point, that is, it can  be removed by a coordinate transformation.

By considering anti-symmetric tensor field ${\tilde b}_{\mu \nu}$ of the action (\ref{18}), one quickly finds that the only non-zero component of ${\tilde H}$
is ${\tilde H}_{{\tilde y}{\tilde u}{\tilde v}} = -\frac{e^{-\tilde v}-4}{2(e^{-\tilde v}-2)^2}$. Consequently, the only non-zero component of
$({\tilde H}^2)_{\mu \nu}$ is $2 {\tilde H}_{{\tilde v}{\tilde y}{\tilde u}} {\tilde H}^{{\tilde y}{\tilde u}}_{\;\;~{\tilde v}}  = -\frac{1}{2}\Big(\frac{e^{-\tilde v}-4}{e^{-\tilde v}-2}\Big)^2$ and
${\tilde H}^2 =0$, too.
Inserting the above results in equations (\ref{18.1}), the conformal invariance conditions up to one-loop (the vanishing of the one-loop $B$-functions) are satisfied
with the dilaton field
\begin{eqnarray}
{\tilde \Phi(\tilde x^\mu)} ~=~{\tilde \Phi_{_{0}}} + {\tilde \Phi_{_{1}}} {\tilde  v} + \ln (\frac{e^{-\tilde v}}{e^{-\tilde v} -2}),\label{18.4}
\end{eqnarray}
where ${\tilde \Phi_{_{0}}}$ and ${\tilde \Phi_{_{1}}}$ are the  constants of integration.

At the end of this subsection we check an interesting result. The dilaton fields (\ref{18.2}) with ${\cal C}=0$ and (\ref{18.4}) with
${\tilde \Phi_{_{0}}}=\Phi_{_{0}}$, ${\tilde \Phi_{_{1}}}=0$ satisfy the following transformations\footnote{For our example, the sigma model constant matrix  $E^+_0(e)$ has given by (\ref{16}). The background fields $E^+(g)$ and ${\tilde E}^+({\tilde g})$ have explicitly written in  (\ref{ex1}).}
\begin{eqnarray}
\Phi(x^\mu) &=&\Phi_{_{0}}+\ln \big(\det {E^+(g)}\big) - \ln \big(\det {E_0^+(e)}\big),\nonumber\\
{\tilde \Phi(\tilde x^\mu)} &=&\Phi_{_{0}}+\ln \big(\det {\tilde E}^+({\tilde g})\big).\label{18.5}
\end{eqnarray}
The above transformations have been obtained by quantum considerations based on a regularization of a functional
determinant in a path integral formulation of Poisson-Lie duality \cite{tyurin} (see also Ref. \cite{n.mohemmadi}).

\subsubsection{\it \large Conformal invariance of the dual sigma model up to the two-loop  $B$-functions}

The two-loop $B$-functions found by Hull and Townsend  \cite{c.hull} (see also Ref. \cite{Tseyt}) are given by
\begin{eqnarray}
B^{^{g}}_{\mu\nu} &=&-\alpha'\Big[R_{{\mu \nu}}-H_{{\mu \rho \sigma}}
H^{{\rho \sigma}}_{\;\;~\nu}+{\nabla}_\mu
{\nabla}_\nu \Phi\Big] -\frac{1}{2} \alpha'^2 \Big[R_{{\mu \rho \sigma \lambda}} R_{\nu}^{{~\rho\sigma \lambda}}
+2 R_{{\mu \rho\sigma \nu }} (H^2)^{\rho\sigma}\nonumber\\
~~&~~~+&2 R_{{\rho\sigma \lambda(\mu }}H_{\nu)}^{~\lambda \delta} H^{\rho\sigma}_{~~_{\delta}} +\frac{1}{3} ({\nabla}_\mu H_{\rho\sigma\lambda})
({\nabla}_\nu H^{\rho\sigma\lambda}) -({\nabla}_\lambda H_{\rho\sigma \mu})
({\nabla}^\lambda H^{\rho \sigma}_{~~\nu})\nonumber\\
~~&~~~+&2 H_{{\mu \rho \sigma}} H_{{\nu \lambda \delta}} H^{{\eta \delta \sigma}} H_{\eta}^{~~\lambda \rho} -2 H_{{\mu \rho}} ^{~~\sigma}  H_{{\nu \sigma \lambda}} (H^2)^{\lambda\rho} \Big]+
{\cal O}(\alpha'^3),\nonumber\\
B^{^{b}}_{\mu\nu} &=&-\alpha'\Big[-{\nabla}^\lambda H_{{\lambda \mu \nu}} +  {\nabla}^\lambda\Phi'  H_{{\mu \nu \lambda}} \Big]
-\frac{1}{2} \alpha'^2 \Big[2 {\nabla}^\lambda H^{\rho \sigma}_{~~[\nu}R_{_{\mu]} \lambda \rho \sigma} +2 ({\nabla}_\lambda H_{\rho\mu\nu}) (H^2)^{\lambda\rho}\nonumber\\
~~&~~~-& 4 ({\nabla}^\lambda H^{\rho \sigma}_{~~[\nu})H_{_{\mu]} \rho \delta } H_{\lambda \sigma}^{~\;\delta}\Big]
+{\cal O}(\alpha'^3),\nonumber\\
B^{^{\Phi}} &=&-\alpha'\Big[-\frac{1}{2} {\nabla}^2 \Phi' + \frac{1}{2} ({\nabla} \Phi')^2-\frac{1}{3} H_{{\mu \nu \rho}} H^{{\mu \nu \rho}} \Big]
-\frac{1}{2} \alpha'^2 \Big[\frac{1}{4} R_{{\mu \rho \sigma \lambda}} R^{{\mu \rho \sigma \lambda}}\nonumber\\
&~~~-&\frac{1}{3} ({\nabla}_\lambda H_{\mu \nu \rho })
 ({\nabla}^\lambda H^{\mu \nu \rho })
-\frac{1}{2} H^{\mu\nu}_{~~\lambda} H^{\rho \sigma \lambda} R_{ \rho \sigma \mu \nu} -R_{\mu \nu} (H^2)^{\mu \nu} +\frac{3}{2} (H^2)_{\mu \nu} (H^2)^{\mu \nu}\nonumber\\
&~~~+& \frac{5}{6} H_{\mu \nu \rho } H^{\mu}_{~~\sigma \lambda} H^{\nu \sigma}_{~~\delta} H^{\rho \lambda \delta}\Big] +{\cal O}(\alpha'^3)
,\label{19.1}
\end{eqnarray}
where $\Phi' = \Phi + \alpha' q H^2$, $(H^2)^{\mu \nu} = H^{\mu \rho \sigma } H_{\rho \sigma}^{~~\nu}$ and $R_{{\mu \rho \sigma \lambda}}$ is the Riemann tensor field.  We note that round brackets  denote the symmetric part on the indicated indices whereas square brackets
denote the anti-symmetric part.
For the line element (\ref{18.3}) of the dual model one finds that the only non-zero components of ${\tilde R}_{{\mu \nu \rho\sigma}}$ and  ${\tilde R}^{{\mu \nu \rho \sigma}}$ are ${\tilde R}_{{\tilde y} {\tilde v} {\tilde u}  {\tilde v}}= - \frac{e^{-2 \tilde v} -4 e^{-\tilde v}+16 }{4(e^{-\tilde v} -2)^3}$ and
${\tilde R}^{{\tilde x} {\tilde y} {\tilde x}  {\tilde u}}= - \frac{e^{-2 \tilde v} -4 e^{-\tilde v}+16 }{4(e^{-\tilde v} -2)}$, respectively. Furthermore,
we find that the only non-zero components of $(\tilde H^2)^{\mu \nu}$ and ${\nabla}_\lambda {\tilde H}_{{\mu \nu \rho}}$ are $2 {\tilde H}^{{\tilde x} {\tilde y} {\tilde u}} {\tilde H}_{{\tilde y} {\tilde u}}^{~~{\tilde x}}= -\frac{1}{2}\Big(\frac{e^{-\tilde v}-4}{e^{-\tilde v}-2}\Big)^2$ and ${\nabla}_{{\tilde v}} {\tilde H}_{{{\tilde y} {\tilde u} {\tilde v}}} = \frac{e^{- \tilde v}}{(e^{-\tilde v} -2)^3}$. We must also note that since  ${\tilde H^2} = 0$, hence, $ \tilde \Phi' = \tilde \Phi$.  Putting all these together into equations (\ref{19.1}) we conclude that the vanishing of the two-loop  $B$-functions satisfy. Therefore the dual model is conformally invariant up to two-loop order.

%%%%%%%%%%%%%%%%%%%%%%%%%%%%%%%%%%%%%%%%%%%%%%%%%%%%%%%%%%%%%%%%%%%%%%%%%%%%%%%%%%%%%%%%%%%%%%%%%%%%%%%%%%%%%%%%%%%%%%%%
\section{\large Worldsheet boundary
conditions under the  Poisson-Lie T-duality}

In this section we generally consider the worldsheet boundary conditions and their transformation
under the Poisson-Lie T-duality. Then, we study the worldsheet boundary conditions  specially for the $H_4$ WZW model.
Consider a $d$-dimensional
target space with $Dp$-branes, i.e., there are $d-(p +1)$
Dirichlet directions along which the field $x^{i}$
($ i = p + 1, ..., d-1$) is frozen. At any given
point on a $Dp$-brane we can choose local coordinates such that
$x^i$ are the directions normal to the brane and $x^{m}$ ($m = 0,
..., p$) as label Neumann directions are coordinates on the brane. Such a coordinate system is called adapted to the brane \cite{Zwiebach}. With this choice the  Dirichlet
condition takes the familiar form
\begin{eqnarray}
\partial_{\tau} x^{i} = 0,~~~~~~~i = p + 1, ..., d-1.\label{Dbrane1}
\end{eqnarray}
The worldsheet boundary is by definition confined to a $D$-brane.
Since the boundary relates left-moving fields $\partial_{+}
x^{^{\mu}}$ to the right-moving fields $\partial_{-}
x^{^{\mu}}$, one can make a general ansatz for this relation.
The goal is then to find the restrictions on this ansatz arising
from varying the action (\ref{666.1}).
The most general local boundary
condition may be expressed as \cite{CC}
\begin{equation}\label{brane2}
\partial_{-}x^{{\rho}}\;=\;{{\cal
R}^{{\rho}}}_{{\nu}}(x^{\mu})\; \partial_{+}
x^{^{\nu}},
\end{equation}
where ${\cal R}^{^{\rho}}_{{\;\nu}}$ is a locally defined
object which is called the {\it gluing matrix}.
This matrix encodes the information about the Neumann and Dirichlet directions  in its
the eigenvalues and eigenvectors.
We assume that ${\cal R}^{^{\rho}}_{{\;\nu}}$ is  in the form of a $2
\times 2$ block matrix as
\begin{equation}\label{brane3}
{{\cal
R}^{{\rho}}}_{{\nu}}(x^{\mu})\;=\;\left(
\begin{tabular}{c|c}
                 ${{\cal
R}^{^{m}}}_{{n}}$ & $0$ \\
\hline
                 $0$ & ${{\cal
R}^{^{i}}}_{{j}}$ \\
                 \end{tabular} \right),
\end{equation}
where  the submatrices ${{\cal
R}^{^{m}}}_{{n}}$ and ${{\cal R}^{^{i}}}_{{j}}$ are
 Neumann-Neumann and Dirichlet-Dirichlet parts, respectively.
By going to adapted coordinates at a point and by using equations (\ref{Dbrane1}) and (\ref{brane2}) we get ${{\cal
R}^{^{i}}}_{{j}}=-{\delta}^{^{i}}_{~j}$. The Neumann condition ${{\cal
R}^{^{m}}}_{{n}}$ still remains very general.

The boundary conditions mentioned above  preserve conformal invariance at
the boundary. We know that each symmetry corresponds to a
conserved current, obtained by varying the action with respect to
the appropriate field. For the case of conformal invariance, the
corresponding current is the  energy-momentum tensor and is
derived by varying the action (\ref{666.1}) with respect to the metric
$h_{\alpha\beta}$. Its components in lightcone coordinates are \cite{Zwiebach}
\begin{equation}\label{brane6}
T_{\pm\pm}\;=\;
\partial_{\pm} x^{^{\mu}}\;
g_{_{ \mu \nu}}\;\partial_{\pm}x^{^{\nu}}.
\end{equation}
The $T_{++}$ component   is called the
left-moving current, whereas $T_{--}$
is referred to as right-moving current. In the conformally invariant case, energy-momentum conservation requires that the $T_{++}$ and $T_{--}$ components
depend only on $\sigma^+$ and $\sigma^-$, respectively.
To ensure conformal
symmetry on the boundary, we need to impose boundary conditions on
the currents (\ref{brane6}). In general, one can find the boundary
condition for a given current  by using its associated charge.
Applied to the energy-momentum tensor, the result is
\begin{equation}\label{brane7}
T_{++}-T_{--}\;=\;0.
\end{equation}
Now, using the equations (\ref{brane2}), (\ref{brane6}) and (\ref{brane7}) we
find
\begin{equation}\label{brane8}
{{\cal{R}}^\rho_{_{{\;\mu}}}}\;
{ g}_{_{\rho\sigma}}\;{{\cal
R}^{^{\sigma}}}_{{\;\nu}}\;=\;g_{\mu\nu}.
\end{equation}
This condition states that the gluing matrix ${{\cal
R}^{^{\mu}}}_{{\;\nu}}$  preserves the metric $g_{\mu\nu}$.

In the next, we investigate structures on $D$-branes. We begin by defining  a Dirichlet projector
${\cal{Q}}^{\mu}_{\;\nu}$ on the worldsheet boundary,
which projects vectors onto the space normal to the brane. These
vectors (Dirichlet vectors) are eigenvectors of ${{\cal
R}^{\mu}}_{\nu}$ with
eigenvalue $-1$. Hence  we can use it to write the Dirichlet condition (\ref{Dbrane1}) on the desired covariant form
\begin{equation}\label{Dbrane3}
{\cal{Q}}^{\mu}_{\;\nu}~\partial_{\tau} x^{\nu} = 0.
\end{equation}
By contracting equations (\ref{brane2}) and (\ref{Dbrane3}), we then obtain
\begin{equation}\label{brane9}
{\cal{Q}}^{^{\mu}}_{_{\;\rho}}\;{{\cal
R}^{^{\rho}}}_{{\;\nu}}\;=\;{{\cal
R}^{^{\mu}}}_{_{\rho}}\;{\cal{Q}}^{^{\rho}}_{_{\;\nu}}=-{\cal{Q}}^{\mu}_{\;\nu}.
\end{equation}
Similarly, we may define a Neumann projector ${\cal
N}^{\mu}_{\;\nu}$ which projects vectors onto the
tangent space of the brane (vectors tangent to the brane are
eigenvectors of ${{\cal R}^{\mu}}_{\nu}$ with eigenvalue $1$) and is defined
as complementary to ${\cal{Q}}^{\mu}_{\;\nu}$, i.e.,
\begin{equation}\label{brane10}
{\cal N}^{\mu}_{\;\nu}={\delta}^{\mu}_{\;\nu}-{\cal{Q}}^{\mu}_{\;\nu},~~~~~~~~~~~~~~{\cal N}^{\mu}_{\;\rho} ~ {\cal Q}^{\rho}_{\;\nu}=0.
\end{equation}
The Neumann projector satisfies the following conditions \cite{lind1}, \cite{lind2}
\begin{eqnarray}\label{brane12}
{\cal N}^\rho_{_{~\mu}}\;
{\cal E}_{_{{\sigma \rho}}}\;{\cal
N}^{\sigma}_{\;~\nu}- {\cal
N}^\rho_{_{~\mu}}\; {\cal E}_{_{
\rho\sigma}}\;{\cal{N}}^{^{\sigma}}_{_{\;\lambda}}\;{{\cal
R}^{^{\lambda}}}_{{\nu}}&=&0,\\\label{brane12.1}
{{\cal{N}^\mu}_{_{{\rho}}}}\;
{ g}_{_{\mu \nu}}\;{{\cal Q}^{^{\nu}}}_{{\;\sigma}}&=&0,\\
{{\cal{N}^\mu}_{_{{\gamma}}}}\;
{{\cal{N}^\rho}_{_{{\nu}}}}\; {{\cal{N}^\delta}_{_{{[\mu , \rho]}}}}  &=& 0.\label{brane12.111}
\end{eqnarray}
The condition (\ref{brane12}) is a condition on the Neumann-Neumann part of ${{\cal
R}^{^{\mu}}}_{{\;\nu}}$. In fact it states the definition of the $b$-field. In adapted coordinates, for a spacefilling brane (along the Neumann directions)
equation (\ref{brane12}) implies, schematically,\footnote{The superscript ``T'' means transposition of the matrix.} ${\cal
R} = {\cal E}^{-1} ~ {\cal E}^{T}$. The condition (\ref{brane12.1}) implies the
diagonalisation of the metric with respect to the $D$-brane and the latter condition is the integrability condition for ${{\cal{N}^\mu}_{_{{\nu}}}}$ \cite{lind2}.

Before starting the main result of this section, let us first to write down the boundary conditions (\ref{brane2}),
(\ref{brane8}), (\ref{brane9}) and  (\ref{brane12})-(\ref{brane12.111}) in the Lie algebra frame (related to the  model on the Lie group). To this end, we use relations
${\Omega}_{ab} = (R^{-1})^{\mu}_{~a}~ g_{\mu\nu}~(R^{-1})^{\nu}_{~b}$, $~{\cal R}^a_{~b} =R_\mu^{~a}~{\cal R}^\mu_{~\nu}~ (R^{-1})^{\nu}_{~b}$ and similarly
for ${{\cal{Q}^\mu}_{_{{\nu}}}}$ and ${{\cal{N}^\mu}_{_{{\nu}}}}$. Then, we get
\begin{eqnarray}\label{brane13}
R_{-}^a &=& {\cal R}^a_{~b}~R_{+}^b,\\
{\cal R}^c_{~a}~ {\Omega}_{cd}~ {\cal R}^d_{~b}~ &=& {\Omega}_{ab},\label{brane13.1}\\
{\cal Q}^a_{~b}~ {\cal R}^b_{~c} = {\cal R}^a_{~b}~ {\cal Q}^b_{~c} &=& -{\cal Q}^a_{~c},\label{brane13.2}\\
{\cal N}^d_{_{~a}}\; {E_{cd}^+}\;{\cal N}^{c}_{\;~b}-
{\cal N}^d_{_{~a}}\; { E_{ dc}^+}\;{\cal{N}}^{^{c}}_{_{\;e}}\;{{\cal
R}^{e}}_{~b}&=&0,\label{brane13.3}\\
{{\cal{N}}^c_{~a}}\;
{\Omega}_{cd}\;{{\cal Q}^d}_{\;b}&=&0,\label{brane13.4}\\
{{\cal{N}}^c_{~a}}\;
{{\cal{N}}^e_{~b}}\; {{\cal {N}}}^d_{~[c , e]} &=& 0.\label{brane13.5}
\end{eqnarray}
From the relation (\ref{brane13}) it is seen that the object ${\cal R}^a_{~b}$ as a gluing map between currents at the worldsheet boundary maps $R_{+}^a$
to $R_{-}^a$, which are elements of the Lie algebra. As explained above in adapted coordinates, the Dirichlet-Dirichlet block of ${\cal R}$ is $-\delta^i_{~j}$. To obtain the non-zero Neumann-Neumann block of ${\cal R}$ one must use the condition (\ref{brane13.3}). Then ${\cal R}$ takes, schematically, the following form \cite{CC}
\begin{equation}\label{brane14}
{{\cal R}}\;=\;\left(
\begin{tabular}{c|c}
                 $({\cal{N}}^T E^+ {\cal{N}})^{-1}~({\cal{N}}^T E^+ {\cal{N}})^T$ & $0$ \\
\hline
                 $0$ & $-\bid$ \\
                 \end{tabular} \right).
\end{equation}
To continue, we obtain the transformation of the gluing matrix which
 defines how the Poisson-Lie T-duality acts on the sigma model boundary conditions.
In this way, we use the canonical transformation of the Poisson-Lie T-duality transformations found by Sfetsos \cite{sfetsos} and \cite{sfetsos1}. The
canonical transformation  between the pairs of variables $(R_\sigma^a~,~P_a)$ and $\big(({\tilde R}_\sigma)_a~,~{\tilde P}^a)$ is given by \cite{sfetsos1}
\begin{eqnarray}\label{brane15}
R_\sigma^a &=& (\delta^a_{~b} - \Pi^{ac} {\tilde \Pi}_{cb}){\tilde P}^b- \Pi^{ab} ({\tilde R}_\sigma)_b,\\
P_a &=& {\tilde \Pi}_{ab} {\tilde P}^b + ({\tilde R}_\sigma)_a,\label{brane16}
\end{eqnarray}
where
\begin{eqnarray}\label{brane17}
R_\sigma^a &=& \frac{1}{2}  (R_+^a - R_-^a),\\
P_a &=&(R^{-1})^{\mu}_{~a} ~ P_\mu = \frac{1}{2} (E_{ba}^+ R_+^b + E_{ab}^+ R_-^b). \label{brane18}
\end{eqnarray}
Now, one can use equations (\ref{15.1}), (\ref{brane17}) and (\ref{brane18}) to write the canonical transformations (\ref{brane15}) and (\ref{brane16})
as a transformation from $R_\pm$ to ${\tilde R}_\pm$. The resulting map is
\begin{eqnarray}\label{brane19}
({\tilde R}_+)_a &=& ({{{\tilde E}}^{+^{-1}})_{ba}}~ ({E^+_0}^{-1})^{cb}~E_{dc}^+~ R_+^d,\\
({\tilde R}_-)_a &=& - ({{{\tilde E}}^{+^{-1}})_{ab}}~ ({E^+_0}^{-1})^{bc}~E_{cd}^+~ R_-^d. \label{brane20}
\end{eqnarray}
Ultimately, by substituting equation (\ref{brane13}) into (\ref{brane20}) and then by using (\ref{brane19}) we obtain
\begin{eqnarray}\label{brane20.1.1}
({\tilde R}_-)_a &=& {\tilde {\cal R}}_a^{~b} ~ ({\tilde R}_+)_b,
\end{eqnarray}
in which \cite{CC}
\begin{eqnarray}\label{brane21}
{\tilde {\cal R}}_a^{~b} ~=~ - ({{{\tilde E}}^{+^{-1}})_{ac}}~ ({E^+_0}^{-1})^{cd}~E_{de}^+~ {\cal R}^e_{~f}~
({E^+}^{-1})^{hf}~{E^+_0}_{gh}~{\tilde E^+}{^{bg}},
\end{eqnarray}
is  the duality transformation of the gluing matrix.
Note that one can immediately get $det {\tilde {\cal R}} = det (-{\cal R})$. We use relation (\ref{brane21}) for
analyzing the dual branes of the $H_4$ WZW model in the following subsection.
%%%%%%%%%%%%%%%%%%%%%%%%%%%%%%%%%%%%%%%%%%%%%%%%%%%%%%%%%%%%%%%%%%%%%%%%%%%%%%

\subsection{\large Example}
\label{subsection4.1}
In this subsection we study the worldsheet boundary conditions and $D$-branes in conformal Poisson-Lie symmetric sigma models generated by
the Heisenberg Lie group  $H_4$  and its dual pair, i.e., the  Lie group $A_2 \oplus 2A_1$. In this example we analyze the consequences of the gluing matrix duality transformation
(\ref{brane21}). The conformal Poisson-Lie T-dual sigma models built by the Drinfel'd double  $\big(H_4~,~ A_2 \oplus 2A_1\big)$ have been given by relation
(\ref{17}) and (\ref{18}). For these models the background fields $E_{ab}^+(g)$ and ${\tilde E}^+{^{ab}}({\tilde g})$ can be obtained from equations (\ref{15.1})-(\ref{15.13}) and (\ref{16}) as follows:
{\small \begin{eqnarray}\label{ex1}
E_{ab}^+(g)~=~\left( \begin{tabular}{cccc}
              $0$ & $0$ & $y e^x$ & $1$\\
              $0$ & $0$ & $-1$ & $0$ \\
              $- y e^x$ & $-1$ & $0$ & $0$ \\
              $1$ & $0$ & $0$ & $0$\\
                \end{tabular} \right),~~~~~~
{\tilde E}^+{^{ab}}({\tilde g})~=~\left( \begin{tabular}{cccc}
              $0$ & $0$ & $0$ & $1$\\
              $0$ & $0$ & $\frac{1}{e^{-\tilde v}-2}$ & $\frac{{\tilde u}}{e^{-\tilde v}-2}$ \\
              $0$ & $-e^{\tilde v}$ & $0$ & ${\tilde y}$ \\
              $1$ & ${\tilde u}~e^{\tilde v}$ & $\frac{{\tilde y}e^{-\tilde v}}{e^{-\tilde v}-2}$ &
              $\frac{2{\tilde y}{\tilde u}}{e^{-\tilde v}-2}$\\
                \end{tabular} \right).
\end{eqnarray}}
Thus, inserting $E_{ab}^+(g)$ and ${\tilde E}^+{^{ab}}({\tilde g})$ from (\ref{ex1}) together with the $E^+_0(e)$ of relation (\ref{16}) into equation (\ref{brane21}) one can get the dual gluing matrix $\tilde {\cal R}$ for any given original matrix ${\cal R}$. From the form of the matrix $E_{ab}^+(g)$
(relation (\ref{ex1})) it is seen that $1 \times 1$, $2 \times 2$ and $3 \times 3$ submatrices  of $E_{ab}^+(g)$ are not invertible. Therefore according to
relation (\ref{brane14}) the brane does not include one, two and or three Neumann directions. Consequently, we can have two different types of $D$-branes:
$D{(-1)}$ and $D3$; that is, all directions are either Dirichlet or Neumann. In the following we compute the dual gluing matrix for each of these cases.

\vspace{2mm}

{\bf Case~(1):} In this case all directions are Dirichlet; that is, ${\cal Q}^a_{~b} = {\delta}^a_{~b}$ and ${\cal N}^a_{~b} = 0$. Since the numbers of Neumann directions are $p+1$, so  we have, in this case, a $D{(-1)}$-brane. From relation (\ref{brane14}) the corresponding gluing matrix is given by
\begin{eqnarray}\label{ex2}
{\cal R}^a_{~b}~=~\left( \begin{tabular}{cccc}
              $-1$ & $0$ & $0$ & $0$\\
              $0$ & $-1$ & $0$ & $0$ \\
              $0$ & $0$ & $-1$ & $0$ \\
              $0$ & $0$ & $0$ & $-1$\\
                \end{tabular} \right).
\end{eqnarray}
Then, equation (\ref{brane21}) yields the dual gluing matrix
\begin{eqnarray}\label{ex3}
{\tilde {\cal R}}_a^{~b}~=~\left( \begin{tabular}{cccc}
              $1$ & $\frac{2({\tilde u} + y e^x)}{e^{-\tilde v}-2}$ & $2 \tilde y$ &
              $\frac{4 {\tilde y} {\tilde u} +2 y {\tilde y} e^{x-\tilde v} {(3-e^{-\tilde v})}}{e^{-\tilde v} -2}$\\
              $0$ & $\frac{-e^{-\tilde v}}{e^{-\tilde v}-2}$ & $0$ & $\frac{-2 \tilde y e^{-\tilde v}}{e^{-\tilde v} -2}$ \\
              $0$ & $0$ & $2 e^{\tilde v}-1$ & $2(y e^{x-\tilde v} - \tilde u e^{\tilde v} -2y e^x)$ \\
              $0$ & $0$ & $0$ & $1$\\
                \end{tabular} \right).
\end{eqnarray}
It has determinant $det {\tilde {\cal R}} = det (-{\cal R})=1$, so the the dual brane may include the following directions:

\vspace{2mm}

$(1.i)~$  Four Dirichlet directions, i.e., the dual brane is also a $D{(-1)}$-brane.
The dual $D{(-1)}$-brane is nontrivially embedded in the dual manifold, and
the embedding can be found by diagonalizing ${\tilde {\cal R}}$. Then, in this case, the only
solution is ${\tilde {\cal R}} = -\bid$, which happens only for backgrounds $E^+(g)$ and ${\tilde E}^+({\tilde g})$ such that
$ {\wt E}^+ ({\wt E}^{+^T})^{-1} = - (E_0^+){^{-1}} E^+ ({ E}^{+^T})^{-1} E_0^+{^T}$.

\vspace{2mm}

$(1.ii)~$  Two Dirichlet directions and two Neumann  directions, i.e., it is a $D{1}$-brane.

\vspace{2mm}

$(1.iii)~$  Zero Dirichlet directions. This is a $D{3}$-brane whose  embedding in $\tilde G$ is given by ${\tilde {\cal R}}$.
Since it is spacefilling it should satisfy the dual version of equation  (\ref{brane13.3}),
${\tilde {\cal R}} = {{\tilde E}}^{+^{-1}} {{{\tilde E}}^{+^{T}}}$. Thus, relation (\ref{brane21}) reduces to
$ \bid =  (E_0^+){^{-1}}  E^+ ({ E}^{+^T})^{-1} E_0^+{^T}$, implying $\Pi \big( E_0^+ +  E_0^+{^T}\big)=0$,
and hence since the condition $ E_0^+ +  E_0^+{^T}=0$ is equal to a vanishing metric we find  $\det \Pi =0$ \cite{CC}.

{\bf Case~(2):} In this case we have  a spacefilling brane, i.e., a $D{3}$-brane. The corresponding gluing matrix according to equation
 (\ref{brane13.3}) is given by
\begin{eqnarray}\label{ex4}
{{\cal R}}~=~ {{ E}}^{+^{-1}} {{{E}}^{+^{T}}} =  \left( \begin{tabular}{cccc}
              $1$ & $0$ & $0$ & $0$\\
              $-2y e^x$ & $1$ & $0$ & $0$ \\
              $0$ & $0$ & $1$ & $0$ \\
              $0$ & $0$ & $-2y e^x$ & $1$\\
                \end{tabular} \right).
\end{eqnarray}
and from (\ref{brane21}) the dual gluing matrix reads
\begin{eqnarray}\label{ex5}
{\tilde {\cal R}}~ = ~ \left( \begin{tabular}{cccc}
              $-1$ & $\frac{-2 \tilde u}{e^{-\tilde v} -2}$ & ${-2 \tilde y}$ & $\frac{-4 { \tilde y} { \tilde u} }{e^{-\tilde v} -2}$\\
              $0$ & $\frac{e^{-\tilde v}}{e^{-\tilde v} -2}$ & $0$ & $\frac{ 2 { \tilde y} e^{-\tilde v}}{e^{-\tilde v} -2}$ \\
              $0$ & $0$ & $1-2 e^{\tilde v}$ & $2 {\tilde u} e^{\tilde v}$ \\
              $0$ & $0$ & $0$ & $-1$\\
                \end{tabular} \right).
\end{eqnarray}
Its determinant is $det {\tilde {\cal R}} =1$, so the dual brane may include the following directions:

$(2.i)~$  Four Dirichlet directions. In this case  we obtain exactly the reverse situation of case (1.iii), that is, we have ${\tilde {\cal R}} =-1$ and
the D3-brane is dual to a D(-1)-brane provided the $\det {\tilde \Pi} $
on $\tilde G$ vanishes.

\vspace{2mm}

$(2.ii)~$  Two Dirichlet directions and two Neumann  directions, i.e.,  a $D{1}$-brane.

\vspace{2mm}

We note that since $det {\tilde {\cal R}} =1$, the dual brane may include zero Dirichlet directions, i.e., a $D{3}$-brane.
But, on the other hand, since it is spacefilling  it should satisfy the dual version of (\ref{brane13.3}),
${\tilde {\cal R}} = {{\tilde E}}^{+^{-1}} {{{\tilde E}}^{+^{T}}}$. In this situation, equation (\ref{brane21}) would require  $E_0^+ +  E_0^+{^T}=0$
and hence a vanishing metric. Thus, we conclude that D3-branes
are dual either to D1-branes  or  D(-1)-branes provided by
$\det \tilde \Pi =0$, but that D3-branes are never dual to D3-branes.

%%%%%%%%%%%%%%%%%%%%%%%%%%%%%%%%%%%%%%%%%%%%%%%%%%%%%%%%%%%%%%%%%%%%%%%%%%%%%%%%%%%%%%%%%%%%%%%%%%%%%%%%%%%%%%%%%%%%%%%%
\section{\large Concluding remarks}
In the present work, first  we have constructed a WZW model based on the Heisenberg Lie group $H_4$ by choosing a convenient parametrization
of the group. The most interesting feature
of our results is the existence of the  Poisson-Lie symmetry in the $H_4$ WZW model.
We have shown that the Poisson-Lie T-duality relates the
$H_4$ WZW model to a sigma model defined on the dual Lie group ${ A}_2 \oplus 2{ A}_1$.
We moreover explicitly worked out the dual model is conformal up to two-loop order and in this manner we have obtained the general
form of the dilaton field  of the dual model.
We have obtained the gluing matrices for the $H_4$ WZW model and its dual model by using the duality map of the gluing matrix obtained by the canonical transformation description of the Poisson-Lie T-duality transformations. We have shown that there are two different cases of the
worldsheet boundary conditions for the $H_4$ WZW model; all directions are either Dirichlet ($D{(-1)}$-brane) or Neumann ($D{3}$-brane).
Case (1)  refers to a $D{(-1)}$-brane; in this case the dual brane includes four Dirichlet directions ($D{(-1)}$-brane) or two Dirichlet directions ($D{1}$-brane), and/or zero Dirichlet directions ($D{3}$-brane). In case (2) we have  a spacefilling brane, i.e., a $D{3}$-brane; in this case we have shown that D3-branes
are dual either to D1-branes  or  D(-1)-branes provided by
$\det \tilde \Pi =0$, but that D3-branes are never dual to D3-branes.

\vspace{8mm}
%%%%%%%%%%%%%%%%%%%%%%%%%%%%%%%%%%%%%%%%%%%%%%%%%%%%%%%%%%%%%%%%%%%%%%%%%%%%%%%%%%%%%%%%%%%%%%%%%%%%%%%%%%%%%%%%%%%%%%%%%%%%%%%%
{\bf \large Acknowledgments}

\vspace{2mm}

This work has supported by the research vice chancellor of Azarbaijan Shahid Madani University under research project No.217/D/10859.

%%%%%%%%%%%%%%%%%%%%%%%%%%%%%%%%%%%%%%%%%%%%%%%%%%%%%%%%%%%%%%%%%%%%%%%%%%%%%%%%%%%%%%%%%%%%%%%%%%%%%%%%%%%

\end{document}